\documentclass[aps,pra,twocolumn,amssymb,superscriptaddress]{revtex4-1}
\usepackage{amsmath}	
\usepackage{graphicx}	
\usepackage{color}
\usepackage{gensymb}

\usepackage{url}
\usepackage{multirow}

\usepackage{hyperref}
\usepackage{upgreek}

\usepackage{ulem}
\newcommand\redsout{\bgroup\markoverwith{\textcolor{red}{\rule[0.5ex]{2pt}{0.4pt}}}\ULon}

\usepackage{fancyhdr}
\usepackage[usenames,dvipsnames]{xcolor}

\begin{document}

\title{Ground-state phase diagram of an anisotropic spin-$1/2$ model on the triangular lattice}

\author{Qiang Luo}
\affiliation{Department of Physics, Renmin University of China, Beijing 100872, China}
\author{Shijie Hu}
\affiliation{Department of Physics and Research Center Optimas, Technical University Kaiserslautern, 67663 Kaiserslautern, Germany}
\author{Bin Xi}
\affiliation{College of Physics Science and Technology, Yangzhou University, Yangzhou 225002, China}
\author{Jize Zhao}
\email{jizezhao@gmail.com}
\affiliation{Institute of Applied Physics and Computational Mathematics, Beijing 100088, China}
\author{Xiaoqun Wang}
\email[]{xiaoqunwang@sjtu.edu.cn}
\affiliation{Department of Physics and Astronomy, Shanghai Jiao Tong University, Shanghai 200240, China}
\affiliation{Department of Physics, Renmin University of China, Beijing 100872, China}
\affiliation{Collaborative Innovation Center for Advanced Microstructures, Nanjing 210093, China}
\date{\today}

\begin{abstract}
Motivated by the recent experiment on a rare-earth material YbMgGaO$_4$
[Y. Li \textit{et al.}, Phys. Rev. Lett. \textbf{115}, 167203 (2015)],
which found that the ground state of YbMgGaO$_4$ is a quantum spin liquid,
we study the ground-state phase diagram of an anisotropic spin-$1/2$
model that was proposed to describe YbMgGaO$_4$.
Using the density-matrix renormalization group method in combination with the exact diagonalization, we calculate
a variety of physical quantities, including the ground-state energy, the fidelity, the entanglement entropy and spin-spin correlation functions.
Our studies show that in the quantum phase diagram there is a $120^{\circ}$ phase and two distinct stripe phases.
The transitions from the two stripe phases to the $120^{\circ}$ phase are of the first order.
However, the transition between the two stripe phases is not the first order, which is different from its classical counterpart.
Additionally, we find no evidence for a quantum spin liquid in this model.
Our results suggest that additional terms may be also important to model the material YbMgGaO$_4$.
These findings will stimulate further experimental and theoretical works in understanding the quantum spin liquid ground state in YbMgGaO$_4$.
\end{abstract}


\maketitle
\section{Introduction}
Frustrated antiferromagnets are the focus of recent research efforts in correlated systems,
largely motivated by the keen interest in searching for exotic states of matter in materials as well as in
microscopic models\cite{Balents-10}. In all proposed states, the quantum spin liquid (QSL)\cite{Anderson-73,Anderson-87}, in particular,
is quite attractive because it is in close association with topological order and can host fractionalized excitations\cite{SB-2016,ZKN-2016}.

Frustration is usually illustrated\cite{Balents-10} by the triangular lattice,
in which the energy of all the bonds can not be simultaneously minimized.
More than forty years ago, Anderson proposed that the ground state of the spin-1/2
antiferromagnetic Heisenberg model on the triangular lattice was a candidate for QSL\cite{Anderson-73,Anderson-87}.
However, extensive numerical calculations have provided strong evidence that
its ground state has a magnetic long-range order with a {${120^\circ}$} structure\cite{HuseElser-88,SLL-94}.
The effect of the frustration only reduces the magnitude of the magnetic order\cite{BLLP-94} and recent density-matrix renormalization
group (DMRG) calculations  show that the magnetization is approximately $M\approx0.205(15)$\cite{WC-07}.

To destroy the magnetic order, one natural way is to include next-nearest-neighbor
interactions, such as the $J_1$-$J_2$ model on the triangular lattice, which has been intensively studied
by coupled cluster method\cite{LBC-15}, DMRG\cite{ZW-15,HGZS-15} and variational Monte Carlo method\cite{ZMQ-15,IHTPB-16,HGS-16} quite recently.
Other proposals are to consider the anisotropic interactions with $J$ along the horizontal direction
and $J'$ along the zigzag direction\cite{WW-11,TS-14}, or the totally random nearest-neighbor interactions\cite{WKNS-14}.

Interest in the triangular lattice is also stimulated by the synthesis of several promising candidate
materials for QSL, which makes it possible to test theoretical predictions experimentally.
These materials, including the inorganic Cs$_2$CuCl$_4$\cite{CTTT-01}, Cs$_2$CuBr$_4$\cite{OTKIMG-03}, and Ba$_3$CoSb$_2$O$_9$\cite{STMK-12}, the organic
salts $\kappa$-(ET)$_2$Cu$_2$(CN)$_3$\cite{SMKMS-03} and EtMe$_3$Sb[Pd(dmit)$_2$]$_2$\cite{Yamashitaetal-10}, have witnessed
the successful composition of the ideal triangular lattice.
Very recently, another triangular lattice material, YbMgGaO$_4$\cite{QMZhang-15,QMZhang-15v2}, was found experimentally to be a
strong candidate for QSL. In this material, ${\rm{Yb}^{3+}}$ sits on a perfect triangular lattice.
It contains thirteen electrons in the $4f$ shell, which shall form spin-orbit entangled Kramers doublets.
These Kramers doublets are split by the D$_{3d}$ crystal fields, and thus can be treated as an effective spin-$1/2$
degree of freedom at low temperature.
Contrary to previous QSL candidate materials, the spin-orbit coupling (SOC) is strong in YbMgGaO$_4$. It is argued that
such SOC leads to anisotropic exchange interactions and eventually destroys the long-range magnetic order.

The rest of the paper is outlined as follows. In Sec. \ref{SEC-ModelvsNM}, we introduce the model Hamiltonian proposed for
YbMgGaO$_4$. In Sec. \ref{SEC-CLPhaseDiagram} the classical phases are obtained by Luttinger-Tisza method.
In Sec. \ref{SEC-QTPhaseDiagram} we provide our phase diagram obtained by the exact diagonalization (ED) method and DMRG method.
The criticality is also discussed.
Sec. \ref{SEC-conc} is devoted to the conclusion,
where we discuss our numerical results as well as the validity of the model Hamiltonian for YbMgGaO$_4$.

\section{Model Hamiltonian}\label{SEC-ModelvsNM}
The model considered in this paper is a highly \textit{anisotropic} spin-1/2 Hamiltonian with nearest-neighbor interactions,
which is proposed to describe YbMgGaO$_4$. It stems from the study of the pyrochlore
lattice\cite{{HFB-04,BIDK-08,SSPPF-12,HCH-14}}, a three-dimensional network of corner sharing tetrahedra,
which offers outstanding opportunities for the study of geometric magnetic
frustration where exotic states such as spin ice can emerge\cite{SB-12,LOB-12,SB-13}.
It should be noted that
whether such an anisotropic exchange on triangular lattice resulted from SOC will stabilize or destabilize the conventional order
is unclear a priori\cite{Nishimoto-2016}.

\begin{figure}[!ht]
\includegraphics[width=7.0cm, clip]{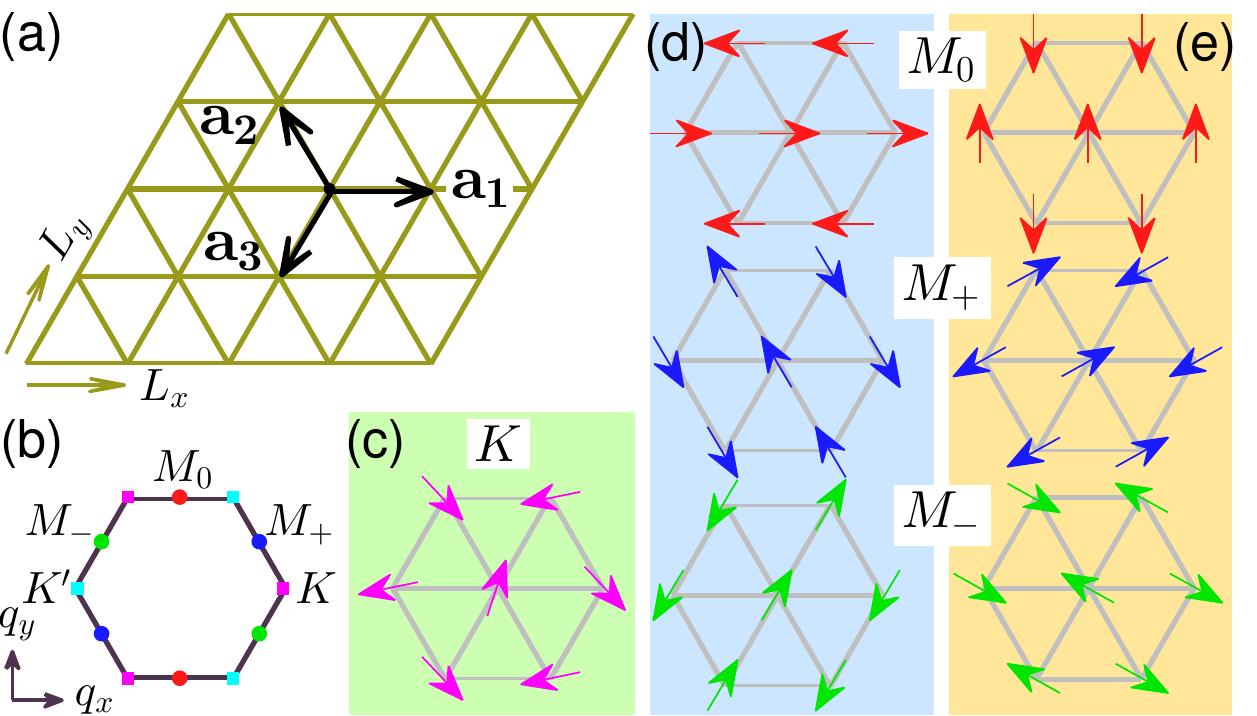}
\caption{(Color online)
(a) Illustration of the anisotropic triangular lattice.
The phase factor $\gamma_{ij}=1$, $e^{i2\pi/3}$ and $e^{-i2\pi/3}$ for the bond along
the $\bf{a}_1$, $\bf{a}_2$ and $\bf{a}_3$, respectively.
(b) The first Brillouin zone of the triangular lattice. The high symmetry points
$K=\big(\frac{4\pi}{3},0\big)$~(magenta) and $K'=\big(-\frac{4\pi}{3},0\big)$~(cyan) at the corners,
and $M_0=\big(0,\frac{2\pi}{\sqrt3}\big)$~(red), $M_+=\big(\pi,\frac{\pi}{\sqrt3}\big)$~(blue) and
$M_-=\big(-\pi,\frac{\pi}{\sqrt3}\big)$~(green) at the middle of the edges are marked.
(c)-(e) show the magnetic patterns of the classical spins.
(c) shows the $120^{\circ}$ order whose peaks of the static structure factors\cite{Notes-SSF} locate at the $K$~($K'$) point.
(d) and (e) show the three degenerate ground states of the stripe-B and stripe-A order, respectively.
The peaks of the static structure factors of the three degenerate states locate at the $M_0$, $M_+$ and $M_-$ points, from top to bottom, respectively.}
\label{FIG-Model}
\end{figure}

In general, the Hamiltonian is given by\cite{QMZhang-15}
\begin{align}
\mathcal{H} =& \sum_{\langle ij\rangle} \bigg[{J_{zz}S_i^zS_j^z+J_{\pm}(S_i^+S_j^- + S_i^-S_j^+)} \nonumber\\
             & +{J_{\pm\pm}(\gamma_{ij}S_i^+S_j^+ + \gamma_{ij}^{*}S_i^-S_j^-)}  \nonumber\\
             & -{\frac{iJ_{z\pm}}{2}(\gamma_{ij}^{*}S_i^+ S_j^z - \gamma_{ij}S_i^-S_j^z + \langle i \leftrightarrow j\rangle)}\bigg]
\label{QSL-Ham}
\end{align}
where $S_{i}^{\alpha}$ ($\alpha=x,y,z$) are the three components of spin-1/2 operators, and $S_{i}^{\pm}=S_i^x \pm iS_i^y$.
The coupling $J_{zz}$ and $J_{\pm}$ are positive in our model.
The phase factor $\gamma_{ij}=1,~e^{i2\pi/3},~e^{-i2\pi/3}$ for the bond $\langle{ij}\rangle$ along the $\bf{a}_1$,
$\bf{a}_2$, $\bf{a}_3$ lattice direction, respectively, see Fig. \ref{FIG-Model} (a).
In the absence of $J_{\pm\pm}$ and $J_{z\pm}$, Eq. \eqref{QSL-Ham} is an XXZ model whose ground
state is known to be the $120^{\circ}$ phase\cite{YMD-14,SZE-15,GMMSOT-15}.
Due to the competition from the $J_{\pm\pm}$ and $J_{z\pm}$,
the ground-state phase diagram is expected to be much richer.
For simplicity, we will set $J_{zz}=1$ as the energy unit. Moreover,
we set $J_{\pm}/J_{zz}=1$ throughout the paper, which agrees with
the recent experiment\cite{Notes-JpmJzpm}.

\section{Classical phase diagram}\label{SEC-CLPhaseDiagram}
We firstly give glimpses of the classical phase diagram of the model \eqref{QSL-Ham} before
moving to the large-scale numerical calculations. The classical phase diagram has already been obtained
by Li \textit{et al}. with Luttinger-Tisza method\cite{LT-46,Bertaut-61,Litvin-74} and Monte Carlo simulation\cite{Metropolis-1953,LWC-2016}.
Here, we will focus on the criticality of the phase transitions between those phases in the phase diagram.
The classical spin is an $O(3)$ vector, which is given by
\begin{equation}\label{classicalspin3}
\boldsymbol{S}_i = S\left(\sin{\theta_i}\cos{\phi_i}, \sin{\theta_i}\sin{\phi_i}, \cos{\theta_i}\right),
\end{equation}
where $\theta_i$ and $\phi_i=\textbf{Q}\cdot\boldsymbol{R}_i+\varphi$ are respectively the polar and azimuthal angles
at site $i$ with the position $\boldsymbol{R}_i$.
The ordering wave vector $\textbf{Q}$ is determined after minimizing $\mathcal{H}$ with respect to $\{\theta_i, \phi_i\}$.
By the Fourier transformation $S_{i}^{\alpha}=\frac{1}{\sqrt{N}}\sum_{\bf{q}}e^{i{\bf{q}}\cdot{\boldsymbol{R}}_i}S_{\bf{q}}^{\alpha}$
with $\alpha=x,~y$ and $z$, the model \eqref{QSL-Ham} takes the form
\begin{eqnarray}
\mathcal{H} & = & \sum_{\left<ij\right>} \sum_{\alpha\beta}S_i^{\alpha}J_{ij}^{\alpha\beta}S_j^{\beta} \nonumber \\
            & = & \sum_{\alpha\beta}\sum_{{\bf{q}}} S_{\bf{q}}^{\alpha}J^{\alpha\beta}(\bf{q})S_{\bf{-q}}^{\beta}
\end{eqnarray}
where $J(\bf{q})$ is a $3\times3$ symmetric matrix, which is written as
\begin{align}\label{JqMatrix}
&J({\bf{q}})=
\left[
    \begin{array}{ccc}
        2J_{\pm}\mathcal{F}+2J_{\pm\pm}\mathcal{G}  & -2\sqrt{3}J_{\pm\pm}\mathcal{K}                & -\sqrt{3}J_{z\pm}\mathcal{K} \\
        -2\sqrt{3}J_{\pm\pm}\mathcal{K}            & 2J_{\pm}\mathcal{F}-2J_{\pm\pm}\mathcal{G}      & J_{z\pm}\mathcal{G}          \\
        -\sqrt{3}J_{z\pm}\mathcal{K}               & J_{z\pm}\mathcal{G}                            & J_{zz}\mathcal{F}            \\
    \end{array}
\right]
\end{align}
with
\begin{align}
\left\{
  \begin{array}{l}
    \mathcal{F} = f({\bf{q}}) = \cos q_x + 2\cos\frac{q_x}{2}\cos\frac{\sqrt{3}q_y}{2} \\
    \mathcal{G} = g({\bf{q}}) = \cos q_x -  \cos\frac{q_x}{2}\cos\frac{\sqrt{3}q_y}{2} \\
    \mathcal{K} = h({\bf{q}}) = \sin\frac{q_x}{2}\sin\frac{\sqrt{3}q_y}{2} \\
  \end{array}
\right..
\end{align}

The smallest eigenvalue of $J({\bf{q}})$ over the first Brillouin zone~(FBZ)~(Fig. \ref{FIG-Model} (b)) provides a lower bound for
the classical ground-state energy\cite{LT-46,Bertaut-61,Litvin-74}. Therefore, we obtain the magnetic order with
the characteristic $\textbf{Q}$ for the given parameters.

\begin{table*}

    \caption{The classical phases, and the characteristic wave vectors $\textbf{Q}$, classical ground-state energy $E_{\textrm{cl}}/(NS^2)$, the allowed angles $\left(\theta,\phi\right)$, and the conditions for the phases for $J_{z\pm}\geq0$.\label{TabPhaseEnergyJzpmGEQ0}} 

    \begin{ruledtabular}

    \begin{tabular}{lllll}

        phases &  $\textbf{Q}$  &  $E_{\text{cl}}/(NS^{2})$  &  $\left(\theta,\phi\right)$ & conditions \\

        \hline

        stripe-B

        &

        $M_{0},M_{\pm}$

        &

        $-2(J_{\pm}-2J_{\pm\pm})$

        &

        $
        \begin{array}{l}
                \left(\frac{\pi}{2},\frac{n\pi}{3}\right)\\
                (n=0,1,\cdots,5)
        \end{array}
        $

        &

        $

        \left\{

            \begin{array}{ll}

                J_{\pm\pm} \leq -\frac{J_{\pm}}{4},&\;\; J_{z\pm} \in \Big[0,\sqrt{\frac{J_{\pm}(3J_{\pm}-J_{zz})}{2}}\Big]\\

                J_{z\pm} \leq \sqrt{2J_{\pm\pm}(J_{zz}+4J_{\pm\pm}-2J_{\pm})},&\;\; J_{z\pm} \in \Big[\sqrt{\frac{J_{\pm}(3J_{\pm}-J_{zz})}{2}},\infty\Big)

            \end{array}

        \right.

        $

        \\


        $120^{\circ}$

        &

        $K,K'$

        &

        $-3J_{\pm}$

        &

        $\Big(\frac{\pi}{2},\forall\phi\in[0,2\pi)\Big)$

        &

        $

        \begin{array}{ll}

            -\frac{J_{\pm}}{4} \leq J_{\pm\pm} \leq \frac{J_{\pm}}{4}-\frac{J_{z\pm}^2}{3J_{\pm}-J_{zz}},&\;\;\;\;\;\;\;\;\;\;\;\;\; J_{z\pm} \in \Big[0,\sqrt{\frac{J_{\pm}(3J_{\pm}-J_{zz})}{2}}\Big]

        \end{array}

        $

        \\


        stripe-A

        &

        $M_{0},M_{\pm}$

        &

        $\alpha$

        \footnote{$\; \alpha = -\left[(J_{\pm}+2J_{\pm\pm}) + J_{zz}/2\right] - \sqrt{4J_{z\pm}^2+\left[(J_{\pm}+2J_{\pm\pm})-{J_{zz}}/{2}\right]^2}$.}

        &

        $\beta$

        \footnote{$\; \beta = \left(\frac{\pi}{4} + \frac{1}{2}{\tan}^{-1}\left[\frac{J_{\pm}+2J_{\pm\pm}-J_{zz}/2}{2J_{z\pm}}\right],
        \frac{(2n+1)\pi}{6}\right)~(n=0,1,\cdots,5)$.}

        &

        $

        \left\{

            \begin{array}{ll}

                J_{\pm\pm} \geq \frac{J_{\pm}}{4}-\frac{J_{z\pm}^2}{3J_{\pm}-J_{zz}},&\;\; J_{z\pm} \in \Big[0,\sqrt{\frac{J_{\pm}(3J_{\pm}-J_{zz})}{2}}\Big]  \\

                J_{z\pm} \geq \sqrt{2J_{\pm\pm}(J_{zz}+4J_{\pm\pm}-2J_{\pm})},&\;\; J_{z\pm} \in \Big[\sqrt{\frac{J_{\pm}(3J_{\pm}-J_{zz})}{2}},\infty\Big)

            \end{array}

        \right.

        $

        \\


    \end{tabular}

    \end{ruledtabular}

    \protect\label{tbl:ecl}

\end{table*}


In table \ref{TabPhaseEnergyJzpmGEQ0}, we present our results for $J_{z\pm}\geq0$.
These results are obtained with the toroidal boundary condition (TBC).
We find three phases in the classical phase diagram, a $120^\circ$ phase\cite{KB-2010} with the magnetic pattern shown in Fig. \ref{FIG-Model} (c),
and two stripe phases\cite{SBK-2012} which are called bond stripe~(stripe-B) phase (Fig. \ref{FIG-Model} (d)) and angle stripe~(stripe-A) phase (Fig. \ref{FIG-Model} (e)).
In the $120^\circ$ phase, the spins lie in the $x$-$y$ plane, leaving the $z$ component disordered.
The peaks of the static structure factors locate at the $K$ and $K'$ and other symmetry equivalent points.
In the stripe-B~(A) phase the spins can be parallel (perpendicular) to one of the three bonds $\bf{a}_1$, $\bf{a}_2$ and $\bf{a}_3$.
Therefore, both of the two stripe phases are three-fold degenerate,
and in the ground state the $\bf{Q}$ locates at one of the three points $M_0$, $M_+$ and $M_-$.
By comparing the ground-state energy in these phases, we can determine the transition points among these phases.

\begin{figure}[!ht]
\includegraphics[width=7.5cm, clip]{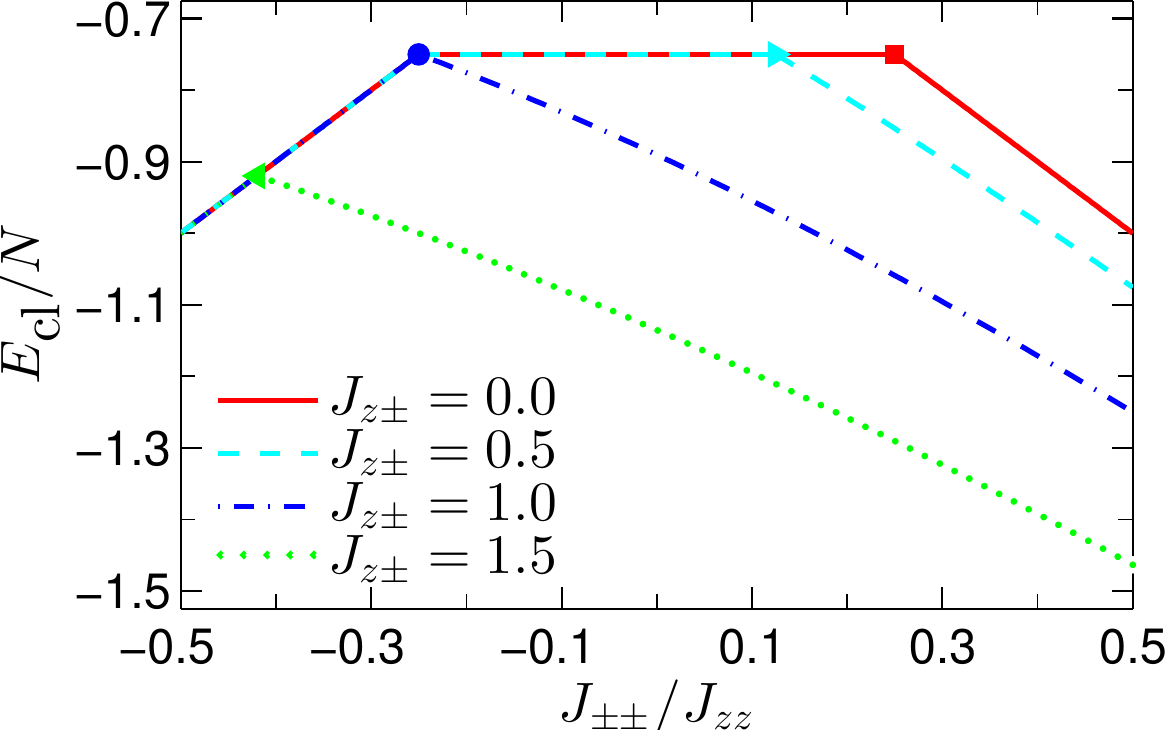}
	\caption{(Color online) Classical ground-state energy per site $E_{\rm{cl}}/N$
	for $J_{z\pm} = $\;0.0 (red solid), 0.5 (cyan dashed), 1.0 (blue dot-dashed)
    and 1.5 (green dotted) with $S=1/2$.
    The filled symbols mark the phase transition points.}\label{FIG-ClGSEnergy}
\end{figure}
FIG. \ref{FIG-ClGSEnergy} shows the ground-state energy for $J_{z\pm} = $\;0.0, 0.5, 1.0 and 1.5.
The symbols in the curves represent the phase transition points.
At $J_{z\pm}=0.0$, we find a $120^{\circ}$ phase sandwiched by two stripe phases.
The inversion symmetry of the curve for $J_{z\pm}=0.0$ with respect to $J_{\pm\pm}=0$ is a reminiscence
of the invariance of Hamiltonian \eqref{QSL-Ham} under the $\pi/2$ rotation around the $z$ axis. This symmetry is broken
for a nonzero $J_{z\pm}$, as can be seen from the curve for $J_{z\pm}=0.5$. The kink in the curve is the
evidence of the first-order phase transition between each stripe phase and $120^{\circ}$ phase.
As $J_{z\pm}$ increases, the transverse components ($x,y$) and the longitudinal component ($z$) become strongly coupled and
the region of $120^\circ$ phase shrinks. Eventually, the $120^{\circ}$ phase vanishes
at the tricritical point $(J_{\pm\pm},J_{z\pm})=(-0.25,1.0)$. The two stripe phases then transit into each other
via a first-order phase transition.

\section{Quantum Phase Diagram}\label{SEC-QTPhaseDiagram}
\subsection{Energy derivative, fidelity and entanglement entropy}\label{SEC-FidvsEE}
In this subsection, we turn to study the quantum phases in the Hamiltonian (\ref{QSL-Ham}),
which may help us to gain some insight into the QSL state in the YbMgGaO$_4$.
First, for a simple profile of the ground-state phase diagram, we study the Hamiltonian (\ref{QSL-Ham}) by
using ED on a $6\times{4}$ cluster with TBC.
We will examine the ground-state energy and its derivative. 
They can provide direct evidences for quantum phase transitions\cite{Sachdev-2011}.
Moreover, we study the ground-state fidelity\cite{VZ-2007,YLG-2007,ZB-2008,Gu-2010,TS-14-Fidelity} and entanglement entropy\cite{OAFF-2002,WSL-04},
which are frequently used as probes for quantum phase transitions in a variety of models.
For a given Hamiltonian with a control parameter $\lambda$ and a ground state $|\psi(\lambda)\rangle$,
the fidelity $F(\lambda,\lambda')$ is defined as the overlap of two wave functions,
\textit{i.e.} $F(\lambda,\lambda')=|\langle\psi(\lambda)|\psi(\lambda^\prime)\rangle|$.
To determine the phase boundary, we choose $\lambda^\prime=\lambda+\delta$, with $\delta\ll{1}$.
The fidelity is expected smaller if $\lambda$ and $\lambda^\prime$ are in different phases than in the same phase.
Therefore, the fidelity shows a minimum around the critical point in our finite systems.
The entanglement entropy, in our case the von Neumann entropy,
is defined as $S_{\textrm{vN}}=-\textrm{Tr}{\rho\ln{\rho}}$, where $\rho$ is the reduced density matrix.
 To calculate $\rho$, we split the system into two halves. The reduced density matrix is obtained by tracing out the freedom of one half.
At the transition point, $S_{\textrm{vN}}$ shows a maximum if such transition is continuous or a jump if the transition is first order\cite{WSL-04}.
Both the fidelity and the entanglement entropy can efficiently determine the phase boundary without the detailed knowledge of the phases.
\begin{figure}[!ht]
\includegraphics[width=8.5cm, clip]{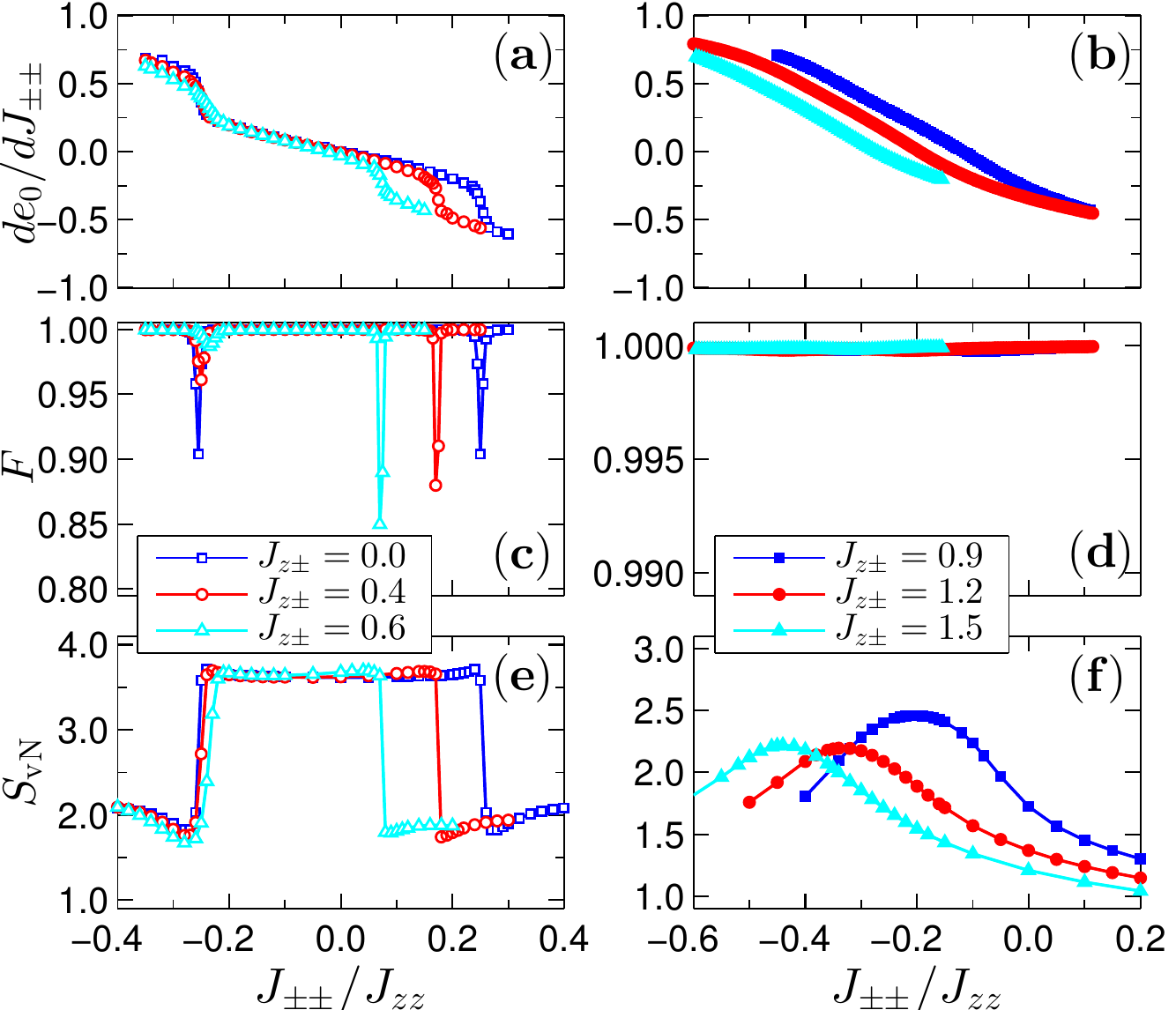}
\caption{(Color online) The first derivative of the ground-state energy ${d e_0}/{d J_{\pm\pm}}$
for (a) $J_{z\pm}=$ 0.0, 0.4 and 0.6 and for (b) $J_{z\pm}=$ 0.9, 1.2 and 1.5 are shown as a function of $J_{\pm\pm}$.
In panel (c) and (d), as well as in panel (e) and (f), the fidelity $F$ as well as the entanglement entropy $S_{\textrm{vN}}$
for the corresponding parameters are shown.
The data are obtained on a $6\times{4}$ clusters.}
\label{FIG-FSvsEE}
\end{figure}

To do so, for a given $J_{z\pm}$, we set $J_{\pm\pm}$ as the control parameter
and take $\delta=0.005$. In Fig. \ref{FIG-FSvsEE}(a),
we show the first derivative of the ground-state energy, ${d e_0}/{d J_{\pm\pm}}$, for $J_{z\pm}=0.0, 0.4$ and $0.6$.
There are two unconspicuous jumps for each curve, which are signals of a first-order phase transition.
One can expect these jumps will become sharp as the system size increases, as shown in the next subsection.
As for the fidelity illustrated in Fig. \ref{FIG-FSvsEE}(c), two dips, which are
interpreted as phase transitions, are seen for each curve of $J_{z\pm}$.
Moreover, as $J_{z\pm}$ increases, the interval between the two dips decreases.
These two transition points merge into one at $(J^0_{\pm\pm}, J^0_{z\pm})=(-0.17(2), 0.85(3))$.
This behavior qualitatively agrees with that in its classical counterpart.
Our results can be further confirmed by the entanglement entropy. In Fig. \ref{FIG-FSvsEE}(e), we show $S_{\textrm{vN}}$ as a function
of $J_{\pm\pm}$.
Two jumps are observed in each curve of $J_{z\pm}$. The positions where the jumps occur agree well with those obtained by the fidelity and energy derivative.
These results suggest a qualitative change in the ground state for our finite clusters and thus a first-order phase transition.

However, when $J_{z\pm}>J^0_{z\pm}$, as shown in panels (b) and (d),
the curves for ${d e_0}/{d J_{\pm\pm}}$ are smooth and
only tiny oscillations (order of $10^{-4}$) are seen in the fidelity $F$.
Therefore, no characteristic behaviors of a first-order phase transition are observed.
Meanwhile, a maximum in $S_{\textrm{vN}}$ is seen in panel (f). These may be taken as a possible signal of a continuous phase transition or a crossover.
\begin{figure}[!ht]
\includegraphics[width=7.0cm, clip]{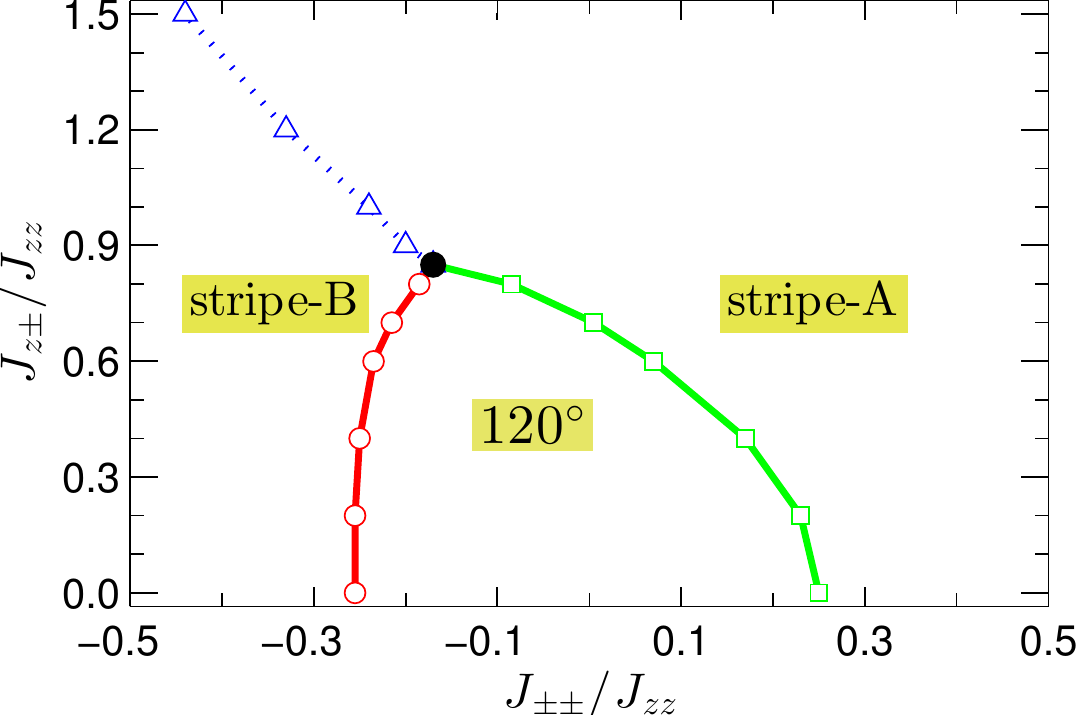}
\caption{(Color online) The schematic quantum phase diagram of the model \eqref{QSL-Ham} on the triangular lattice.
There are three different phases, the $120^{\circ}$ phase,
the stripe-B phase and stripe-A phase. The solid boundary lines represent a first-order phase transition,
while the dotted line separating the stripe-B phase and the stripe-A phase stands for a continuous transition or
a crossover.}
\label{FIG-PDED}
\end{figure}

In Fig. \ref{FIG-PDED}, we summarize our phase diagram. The phase boundary is obtained from ED.
It includes three phases, stripe-B phase, stripe-A phase and $120^\circ$ phase.
The transitions from the two stripe phases to the $120^\circ$ phase are of the first order.
Above the tricritical point, \textit{i.e.} $J_{z\pm}>J_{z\pm}^0$, the $120^\circ$ phase disappears.
The stripe-B phase transits into the stripe-A phase directly as $J_{\pm\pm}$ increases.
Our analysis of the classic spin model tells us that such transition is first order.
However, our ED results do not detect signals for a first-order phase transition.
In addition, we confirm these conclusions on hexagonal clusters as well. 
In the following subsections, we will discuss more about the properties of each phase and phase transitions.

\subsection{Ground-state energy and magnetic structure factors}\label{SEC-SMSF}
In the last subsection, we have mapped out the phase boundary by ED but without providing the details of those phases.
Here, we resort to DMRG\cite{White-9293,PWKH-1999,SchRMP-2005}, which enables us to access large lattice sizes, to provide more information of those phases.
To get accurate results with DMRG, we use cylindrical boundary condition (CBC). The clusters we use are $L_x\times L_y = $$10\times{6}$,
$12\times{8}$ and $15\times{10}$. The aspect ratio is about 1.5, which was proposed to be the best to minimize the edge effect\cite{WC-07,StouWhite-ARCMP-12}.
We keep up to 3000 states in our simulations and typically about 10 sweeps are performed
to improve the accuracy.

\begin{figure}[!ht]
\includegraphics[width=8.0cm, clip]{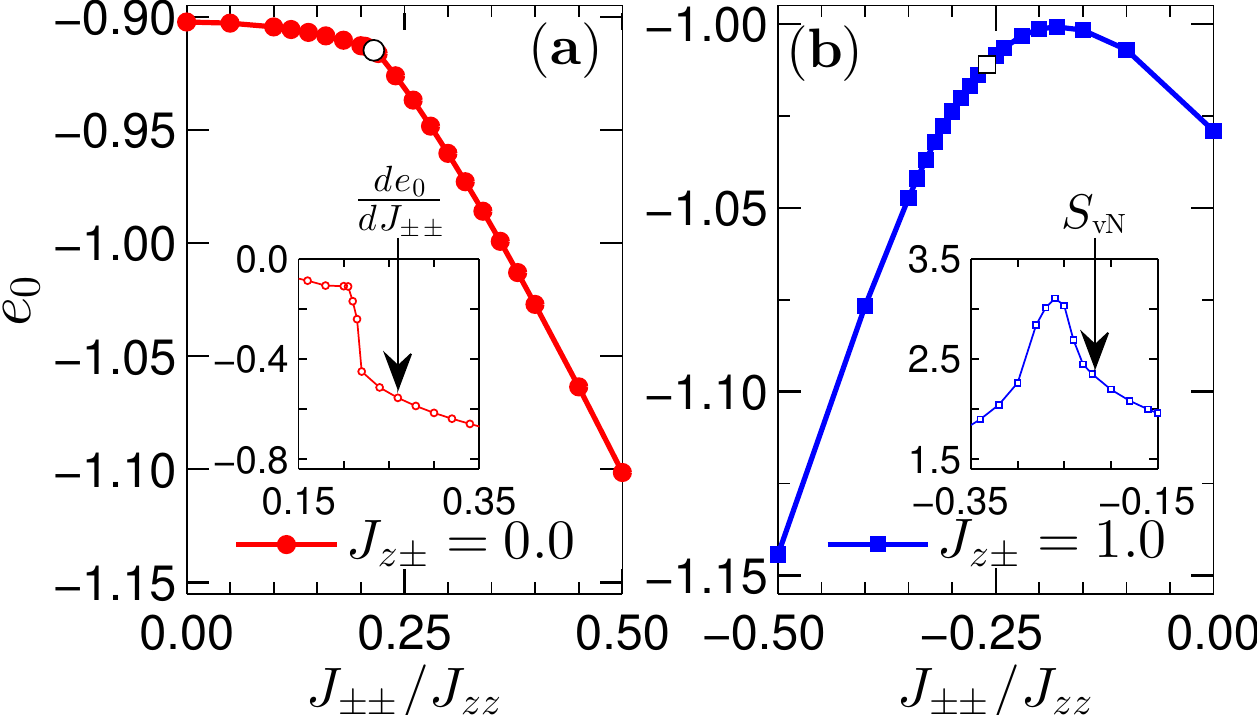}
\caption{(Color online) The ground-state energy per site $e_0$ for $J_{z\pm}=0.0$ (a) and $J_{z\pm}=1.0$ (b) of $10\times6$ cluster.
Inset: (a) A jump in ${d{e_0}}/{d{J_{\pm\pm}}}$ suggests a first-order phase transition, and the position is marked by an open circle ($\circ$)
in the main panel. (b) The position of the maximum in the entanglement entropy $S_{\textrm{vN}}$ is
marked by open square ($\square$) in the main panel.}
\label{FIG-VNEvsGSE}
\end{figure}

Let us start our discussion by showing the ground-state energy. The energy per site $e_0$ as a function of $J_{\pm\pm}$ is shown in FIG. \ref{FIG-VNEvsGSE}.
In panel (a), we show the energy for $J_{z\pm}=0.0$.  A clear kink can be observed in the energy curve.
The position of the kink, $J_{\pm\pm}=0.215(5)$, is marked by an open circle.
It is remarkable to note that such a kink is characteristic of a first-order phase transition.
In contrast, in panel (b) with $J_{z\pm}=1.0$, the energy is smooth as a function of $J_{\pm\pm}$ within our numerical accuracy.
This provides further evidence that such transition is not first order.
The possible phase transition point (the open square) $J_{\pm\pm}\simeq-0.265(5)$ is given by the sharp peak in the entanglement entropy,
as shown in the inset.

Now we turn to study the magnetic order in each phase.
The magnetic order is naturally detected by the spin-spin correlation functions
\begin{eqnarray}\label{SReal}
\mathcal{S}^\nu_{ij}=\langle{S^\nu_i {S^\nu_j}}\rangle
\end{eqnarray}
and their Fourier transformation, \textit{i.e.}, static magnetic structure factors (SMSF),
\begin{eqnarray}\label{SMomentum}
\mathcal{S}_N^\nu({\bf{Q}})=\frac{1}{N}\sum_{ij}e^{i{\bf{Q}}\cdot{({\bf{R}}_i-{\bf{R}}_j)}}\mathcal{S}^\nu_{ij}
\end{eqnarray}
where $\nu=x,\;y$ and $z$ represents the spin component, $\langle\cdots\rangle$ is the average over the ground state,
${\bf{R}}_i$ is the position of site $i$, and $N=L_xL_y$ is the total number of the spins.

\begin{figure}[ht]
\includegraphics[ width=8.5cm, clip]{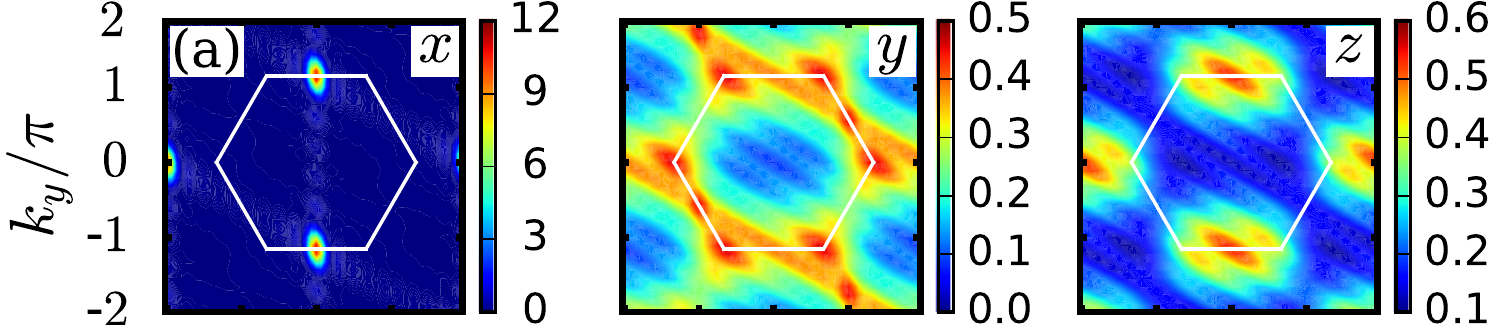}
\includegraphics[ width=8.5cm, clip]{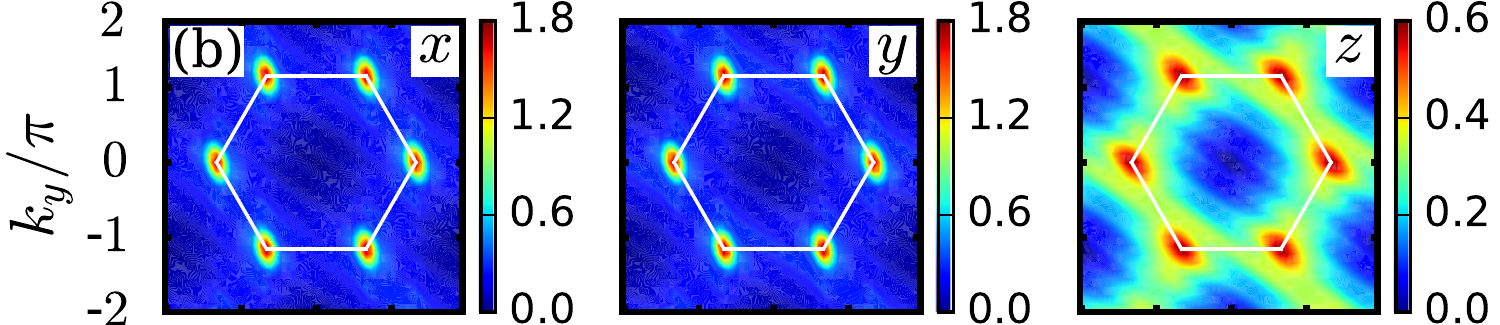}
\includegraphics[ width=8.5cm, clip]{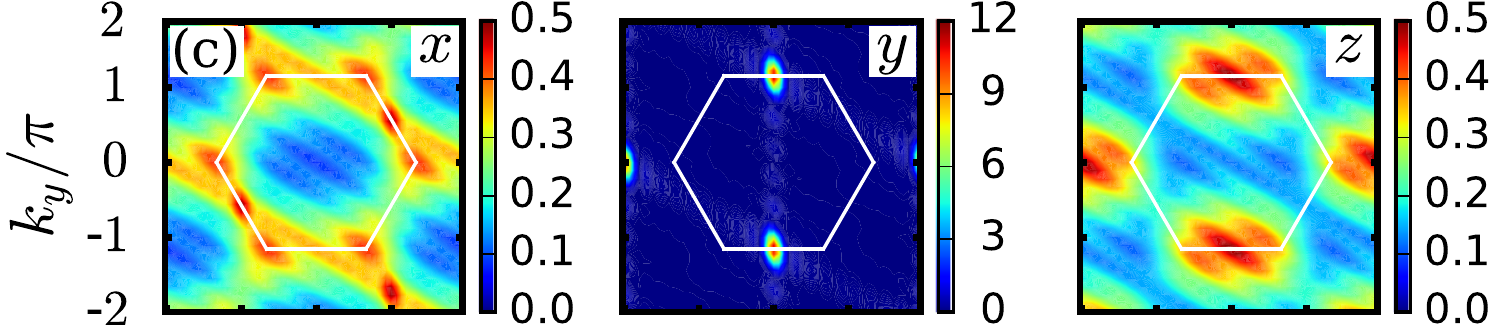}
\includegraphics[ width=8.5cm, clip]{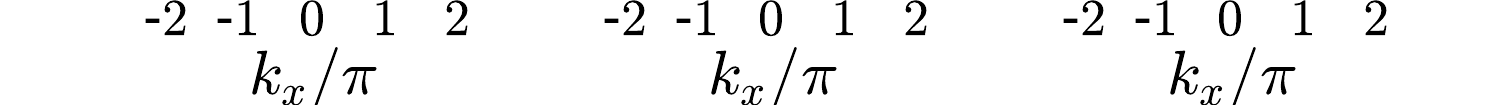}
\caption{(Color online) Typical contour plots of the SMSF of $x$, $y$ and $z$ components
	for $J_{\pm\pm}=-0.38$ (a), $0.10$ (b) and $0.40$ (c) along the line $J_{z\pm}=0$. The white lines denote the FBZ.
        $x,y$ and $z$ represent the spin components.}
\label{FIG-MSF120Stripe}
\end{figure}

In Fig. \ref{FIG-MSF120Stripe}, we show the typical SMSF of the stripe-B phase, $120^\circ$ phase and stripe-A phase at $J_{z\pm}=0.0$.
The positions of the peaks of the SMSF clearly demonstrate the differences among the three phases.
In the stripe-B phase, as we show in row (a) where $J_{\pm\pm}=-0.38$, the dominant spin component is $x$,
and the peak of the SMSF locates at $M_0$.
However, in the stripe-A phase shown in row (c) with $J_{\pm\pm}=0.40$, the dominant spin component becomes $y$
and the peak remains at $M_0$.
Actually, the ground states of the two stripe phases are three-fold degenerate,
and the peak of the SMSF can locate at any of the equivalent point $M_0$, $M_+$ or $M_-$.
However, we do not detect all the degenerate states simultaneously in our DMRG calculations.
This results from the CBC we imposed to simulate the highly anisotropic triangular lattice.
In Appendix \ref{AppA}, we will discuss more about this.
Hereafter, for simplicity, we will restrict our discussion to ${\bf{Q}}_M=M_0=(0,\frac{2\pi}{\sqrt{3}})$ only because others are symmetrically equivalent.
In the $120^\circ$ phase shown in row (b) with $J_{\pm\pm}=0.10$,
both $x$ and $y$ components are dominant and of nearly equal weight.
They peak at ${\bf{Q}}_K=K=(\frac{4\pi}{3},0)$.
The behaviors in the $120^\circ$ phase agree with that in the standard XXZ Heisenberg model\cite{GMMSOT-15}.
From these data, one can immediately figure out that
the magnetic order in the quantum model is similar to its classical version.

As $J_{z\pm}$ increases, the spins in the stripe-B phase remain in the $x$-$y$ plane,
but in the stripe-A phase  the $z$ component also becomes dominant.
This phenomenon can be understood directly from $\mathcal{H}_{z\pm}$,
the last line of the Hamiltonian \eqref{QSL-Ham}, as follows.
In the mean field level, $\mathcal{H}_{z\pm}$ is approximately zero in the stripe-B phase, and thus
$J_{z\pm}$ is irrelevant in the stripe-B phase.
However, in the stripe-A phase, $\mathcal{H}_{z\pm} \sim 2J_{z\pm}\sum_{i}S_i^{z}$.
Therefore, the $z$ component should be ordered for a finite $J_{z\pm}$.
\begin{figure}[ht]
\includegraphics[width=7.0cm, clip]{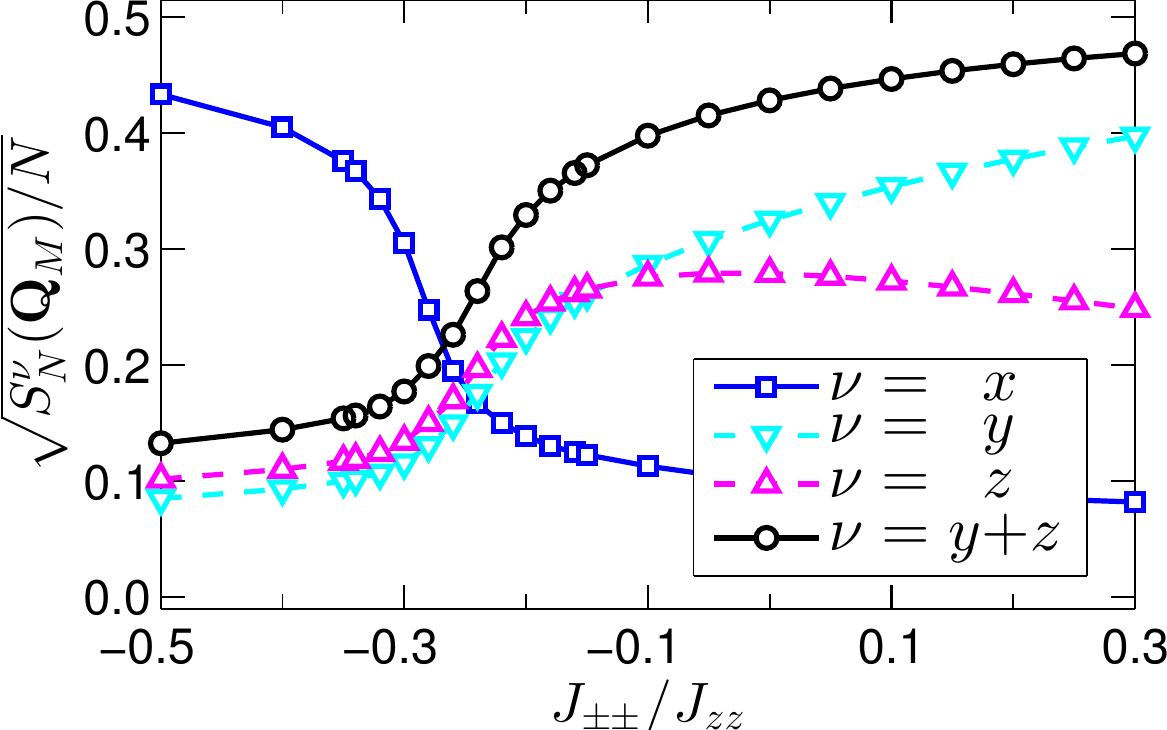}
	\caption{(Color online) The $x$~(blue), $y$~(cyan), $z$~(magenta) and $y+z$~(black) components of the SMSF at ${\bf{Q}}_M$ are shown as a function of $J_{\pm\pm}$ when $J_{z\pm}=1.0$. The size of the cluster is $L_x\times L_y=10\times6$.}
\label{FIG-JZPM0}
\end{figure}

In Fig. \ref{FIG-JZPM0}, we show $\sqrt{\mathcal{S}_N^\nu({\bf{Q}}_M)/N}$ as a function of $J_{\pm\pm}$ for $J_{z\pm}=1.0$,
which is well above the tricritical point and only the two distinct stripe phases exist.
One can distinguish from Fig. \ref{FIG-JZPM0} that the SMSF
in both stripe phases peak at ${\bf{Q}}_M$.
From our previous analysis of the energy derivative, the fidelity, and the entanglement entropy, we do not find signals of a first-order
phase transition between the two stripe phases. This can be further clarified by the order parameters in each phase.
To investigate the phase transition between the two stripe phases,
we make a comparison between the $x$ component and the summation of $y$ and $z$
components of the SMSF, which is also shown in Fig. \ref{FIG-JZPM0}. The intersection of the two curves occurs at
$J_{\pm\pm}=-0.27(1)$, and is fairly consistent with the one obtained by the entanglement entropy in Fig. \ref{FIG-VNEvsGSE}(b).
Moreover, these quantities evolve smoothly as a function of $J_{\pm\pm}$, which provides us another evidence to exclude the possibility
of a first-order phase transition between the two stripe phases. Therefore, there are two possibilities
about the transition between the two phases. One is that the transition is continuous, and the other is a crossover.
However, due to the limited sizes, we can not tell definitely which one is correct and therefore leave it as an open question.

\subsection{Numerical results relevant to YbMgGaO$_{4}$}\label{subSEC-YbMgGaO4}
Since it is argued that the Hamiltonian \eqref{QSL-Ham} is sufficient to describe the nature of YbMgGaO$_{4}$\cite{QMZhang-15, LiChen-arXiv1608},
in this subsection, we will focus on this material with the relevant coupling parameters.
These parameters have been determined accurately by measuring the magnetization and magnetic susceptibility as well as by the
electron spin resonance (ESR)\cite{QMZhang-15}. According to those experiments, $J_{z\pm}$ is rather small and is insignificant in the material,
and this inference is further confirmed by the later experiment on the magnetic excitations\cite{Padd-Yb-2016}.
Though the ESR can only determine the intensity but not the sign of the $J_{\pm\pm}$ term, the sign of the $J_{\pm\pm}$
does not have any effect on the ground-state phases if $J_{z\pm}=0.0$. This is because that in the absence of the $J_{z\pm}$ term,
the Hamiltonian \eqref{QSL-Ham} is invariant under the $\pi/2$ rotation around the $z$ axis. Consequently, in the following discussions,
we will focus on $J_{zz}=J_{\pm}=1.0$, $J_{z\pm}=0.0$, and sweep the parameters in the region  $0.0\le J_{\pm\pm}\le 0.5$ hereafter.
The validity of our conclusions to other $J_{z\pm}$ is also confirmed by our DMRG calculations.

To search for the possible QSL phase and distinguish the phase boundaries, we introduce the order parameter
\begin{eqnarray}
M_N({\bf{Q}})=\sqrt{\sum_{\nu}\mathcal{S}_N^\nu({\bf{Q}})/N}
\end{eqnarray}
where ${\bf{Q}}={\bf{Q}}_K$ or ${\bf{Q}}_M$.
\begin{figure}[!ht]
\includegraphics[width=7.0cm, clip]{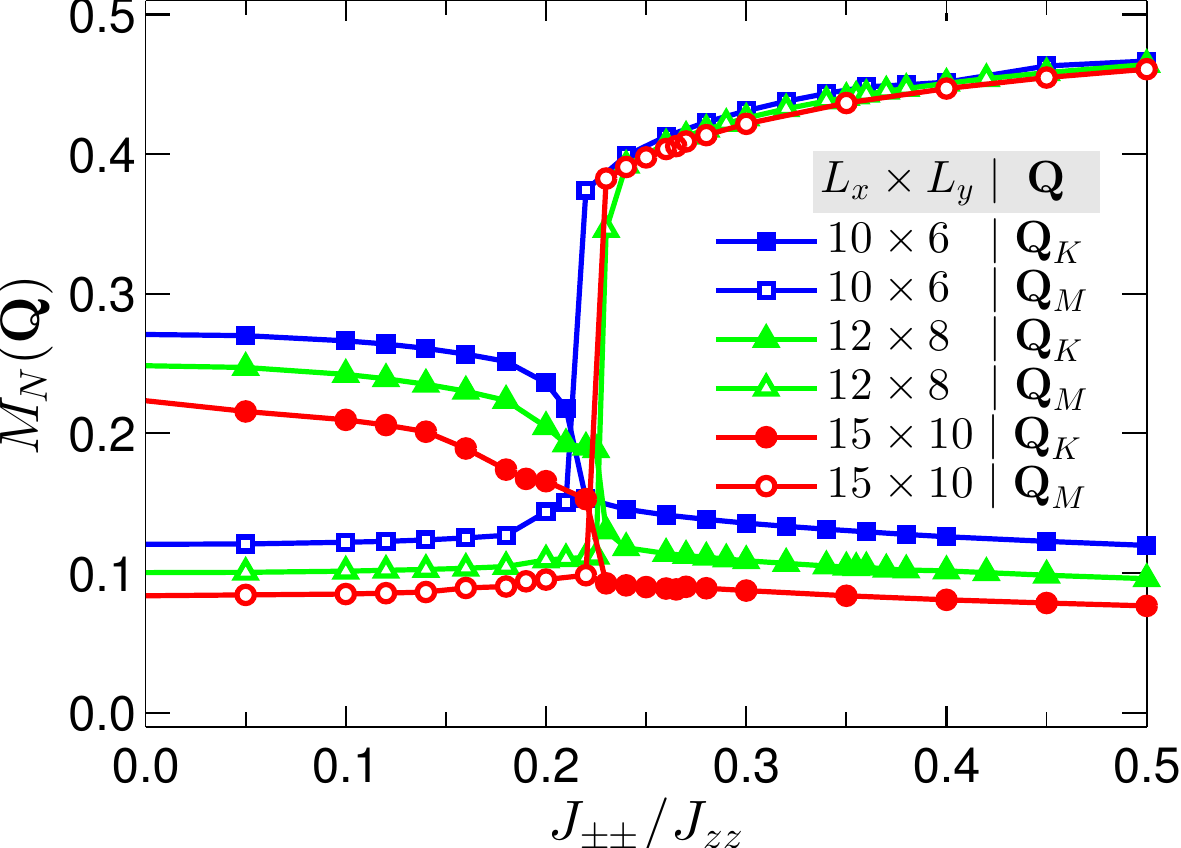}
\caption{(Color online) Order parameters $M_N(\textbf{Q})$ for the $120^{\circ}$ order (the filled symbols)
and stripe order (the open symbols) with $\textbf{Q}={\bf{Q}}_K$ and ${\bf{Q}}_M$, respectively.
The sizes of the clusters are $L_x\times L_y$ = $10\times6$ (blue), $12\times8$ (green) and $15\times10$ (red).}
\label{FIG-MSF}
\end{figure}
In Fig. \ref{FIG-MSF}, we plot $M_N(\bf{Q})$ at ${\bf{Q}}_K$ (filled symbols) and ${\bf{Q}}_M$ (open symbols)
as a function of $J_{\pm\pm}$ for $J_{z\pm}=0.0$ and $L_x\times L_y=10\times6,\;12\times8$ and $15\times10$.
We observe a jump at $J_{\pm\pm}=J^c_{\pm\pm}=0.22(1)$ for all the curves.
This sharp transition indicates a first-order phase transition,
which is fairly in accordance with the independent verdicts in Fig. \ref{FIG-FSvsEE}.
Below $J^c_{\pm\pm}$, it is $120^\circ$ ordered and above $J^c_{\pm\pm}$ the stripe-A order dominates.
Remarkably, in the stripe phase the $M_N({\bf{Q}}_M)$ is nearly independent of the lattice size. Near the transition point,
$M_N(\bf{Q})$ at both ${\bf{Q}}_M$ and ${\bf{Q}}_K$ are slightly suppressed but remains finite.
Therefore, our results exclude the possibility of a QSL ground state in this model.

Reasons for the failure to detect the QSL phase in Eq. \eqref{QSL-Ham} numerically are perplexing.
We conjecture that the Hamiltonian \eqref{QSL-Ham} may be incomplete to describe the nature of the compound YbMgGaO$_4$,
and additional terms should be taken into account\cite{Padd-Yb-2016}.
This is partially because as an spin-orbit-coupled insulator with odd number of electrons per unit cell\cite{QMZhang-15},
its ground state must be exotic if the time-reversal symmetry is not broken according to the recent extension\cite{WPVZ-2015}
of the Hastings-Oshikawa-Lieb-Schultz-Mattis theorem\cite{Hastings-2004,Oshikawa-2000,LSM-1961}.
The classical and semiclassical \cite{LWC-2016,LWY-Yb-2016} analysis make the ground-state phases of the Hamiltonian \eqref{QSL-Ham}
be obscure~(both are against the QSL phase), while various contemporaneous experiments\cite{Shen-Yb-2016,Padd-Yb-2016,Li-Yb-2016,SYLi-16}
coincide with each other, and all show that the ground state of YbMgGaO$_4$ is a strong candidate for a gapless $U(1)$ QSL phase.
Remarkably, measurements of the magnetic excitations close to the field-polarized state indicate that
the next-nearest-neighbor interactions may be indispensable to get the full nature of the YbMgGaO$_4$\cite{Padd-Yb-2016}.

\section{Conclusion}\label{SEC-conc}
In summary, by using ED and DMRG method, we study the ground-state phase diagram of an anisotropic spin-$1/2$ model with nearest-neighbor
anisotropic interactions on the triangular lattice proposed to describe the YbMgGaO$_4$.
We utilize the ground-state energy and its derivative, the fidelity and the entanglement entropy as probes for phase transitions, 
and the magnetic structure factor to distinguish the phases therein. Our numerical results show 
that there are three distinct phases: a $120^{\circ}$ phase with three sublattice sandwiched 
by two stripe phases. Our large-scale DMRG calculations suggest that the $120^{\circ}$ phase 
and the stripe phases in model \eqref{QSL-Ham} are robust enough against the quantum fluctuations.
The effects of the quantum fluctuations merely change the phase boundary compared with its classical counterpart.
Although our result does not favor an intermediate nonmagnetic phase near the classical phase boundaries,
nevertheless it can not exclude the existence of the QSL ground-state phase for YbMgGaO$_4$.
We attribute the possible reason to the incompleteness of the microscopic model \eqref{QSL-Ham}.
It calls for more accurate experiments to settle this issue.
Moreover, in the classical phase diagram, the transition between the two stripe phases is of the first order,
but our numerical results exclude such possibility in the quantum one.

\textit{Note added}. Recently, we became aware of a preprint \cite{ZZhu-17} on a similar topics.

\acknowledgments
We thank G. Chen for enlightening discussions as well as some comments and suggestions on the draft.
We acknowledge C. Liu, R. Yu, Y.-C. He and X.-F. Zhang for some discussions.
S. Hu was supported by the Nachwuchsring of the TU Kaiserslautern,
and by the German Research Foundation (DFG) via the Collaborative Research Centers SFB/TR49,
J. Zhao was supported by the the National Natural Science Foundation of China (Grants No. 11474029).
X. Wang was supported by the National Program on Key Research Project (Grants No. 2016YFA0300501),
and by the National Natural Science Foundation of China (Grant No. 11574200).
We gratefully acknowledge the computing time granted by the John von Neumann Institute for Computing (NIC)
and provided on the supercomputer JURECA at J\"ulich Supercomputing Centre (JSC).
We also thanks the computational resources provided by Physics Laboratory for High Performance Computing~(RUC),
and by Shanghai Supercomputer Center where most of the computations are carried out.

\appendix
\section{Note on the three-fold-degenerate ground states in the stripe phases}\label{AppA}
\setcounter{figure}{0}
\renewcommand{\thefigure}{A.\arabic{figure}}
\setcounter{equation}{0}
\renewcommand{\theequation}{A.\arabic{equation}}

The ground states in both the stripe-B and stripe-A phases are three-fold degenerate \cite{Notes-inversion}
if TBC are used in our simulations. The three states in each phase can be distinguished
by the positions~($M_0$, $M_+$ or $M_-$) of the peaks of the SMSF.
However, to get accurate results, we use CBC instead of TBC in our DMRG simulations,
which lifts the degeneracy of the ground states.
This is because, under CBC, the number of the bonds along three directions $\bf{a}_1$, $\bf{a}_2$ and $\bf{a}_3$ are not equal.
Consequently, the energy of the three states is different, depending on the size of the lattice and the parameters.
As the parameters change, one or two of the three states may have lower energy than the others.

\begin{figure}[!htb]
\includegraphics[width=8.cm, clip]{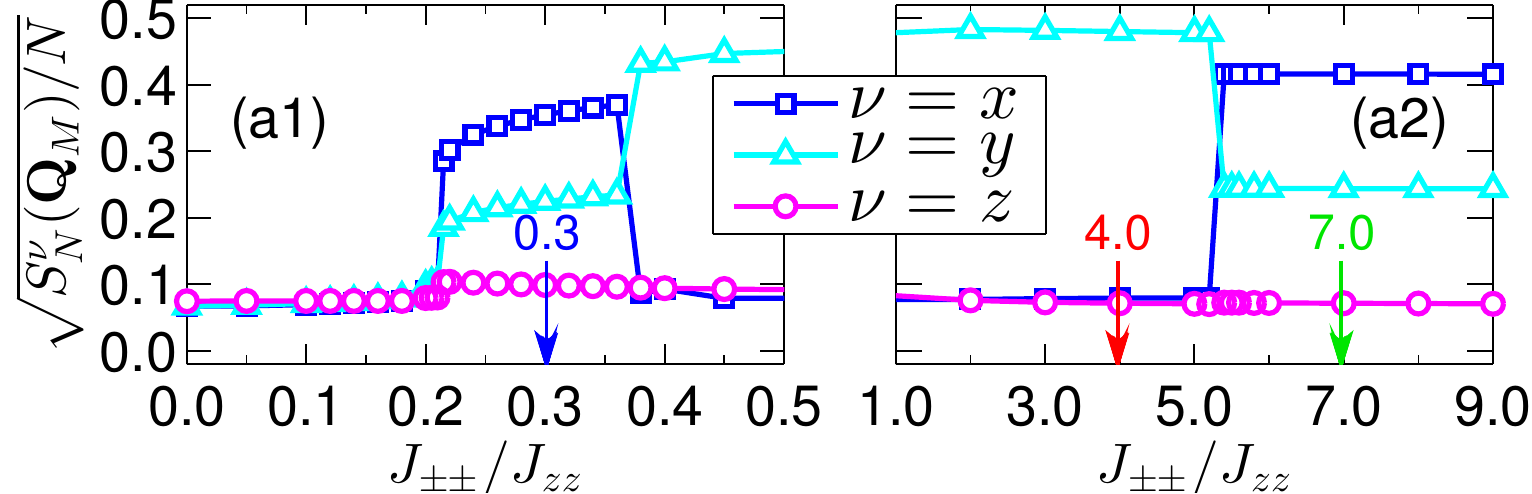}
\includegraphics[width=8.cm, clip]{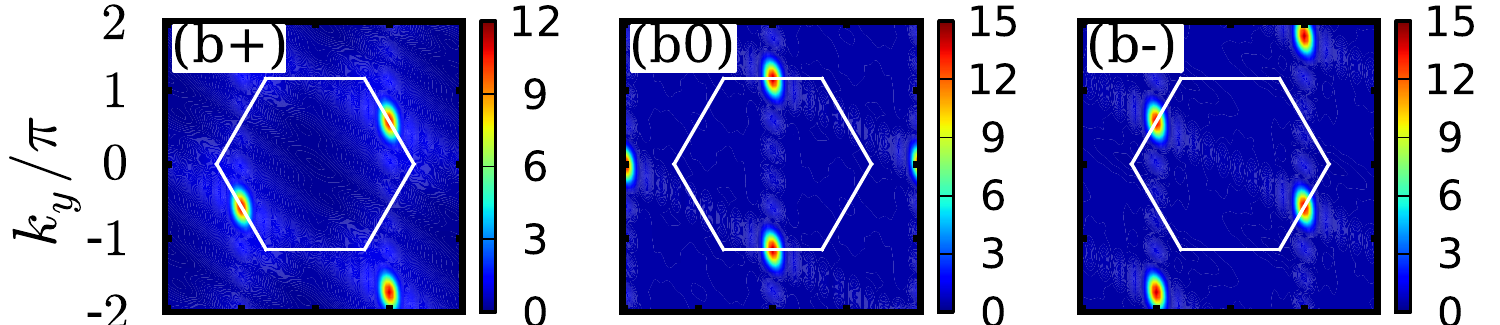}
\includegraphics[width=8.cm, clip]{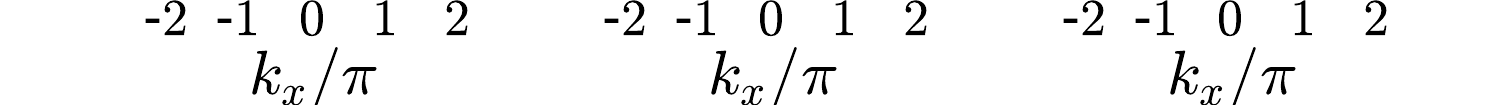}
\caption{(Color online)
Upper panel: the $x$, $y$ and $z$ components of the SMSF at ${\bf{Q}}_M$
for $L_x\times L_y$ = $10\times6$ cluster under CBC are shown as a function of $J_{\pm\pm}$.
Lower panel: (b+), (b0) and (b-) show the contour plots of the SMSF at $J_{\pm\pm}=$0.3, 4.0 and 7.0, respectively.
Here, all three components are summed up.}
\label{FIG-3foldStripe}
\end{figure}

To verify our explanation, we show our numerical results for $L_x\times L_y$ = $10\times6$ clusters under CBC
in FIG. \ref{FIG-3foldStripe}. The $J_{z\pm}$ is set zero for simplicity, and $J_{\pm\pm}$ is a tunable
parameter. We can see that the stripe-A phase emerges when $J_{\pm\pm}\gtrsim0.215(5)$. This phase
is split into three regimes at $J_{\pm\pm}^{c1}\simeq0.37(1)$ and $J_{\pm\pm}^{c2}\simeq5.3(1)$.
In these three regimes, i.e., $J_{\pm\pm}$$<$$J_{\pm\pm}^{c1}$, $J_{\pm\pm}^{c1}$$<$$J_{\pm\pm}$$<$$J_{\pm\pm}^{c2}$, and
$J_{\pm\pm}$$>$$J_{\pm\pm}^{c2}$, the SMSF peaks at $M_+$, $M_0$ and $M_-$, respectively.
The three different positions of the peaks are the evidence of the three-fold-degenerate ground states in the stripe-A phase.
Similarly, one can arrive at the same conclusion in the stripe-B phase.
\bibliography{manuscript}

\begin{thebibliography}{99}
\bibitem{Notes-JpmJzpm} The experimental data of Ref. \cite{QMZhang-15} showed that $J_{\pm}/J_{zz}=0.92(16)$, and $\vert J_{z\pm}\vert/J_{zz}=0.04(8)$.
\bibitem{Notes-SSF} The mathematical definition of the static structure factor will be given bellow in Eq. \eqref{SReal} and \eqref{SMomentum}.
\bibitem{Notes-inversion} Flipping all the spins simultaneously makes no difference in the structure factors so we do not distinguish corresponding degeneracy here.
\end{thebibliography}

\end{document}